\documentclass[referee]{aa} % for a referee version
\usepackage{graphicx}
\usepackage{txfonts}
\newcommand{\mdot}{\mbox{$\dot{M}$}}

\newcommand{\pdot}{\mbox{$\dot{p}$}}
\newcommand{\Cdot}{\mbox{$\dot{C}$}}
\newcommand{\nf}{\mbox{${\cal N}$}}

\def \etal{et~al.\/}

\def\lesssim{\mathrel{\hbox{\rlap{\hbox{\lower4pt\hbox{$\sim$}}}\hbox{$<$}}}}

\def\gtrsim{\mathrel{\hbox{\rlap{\hbox{\lower4pt\hbox{$\sim$}}}\hbox{$>$}}}}

\begin{document}

\title {Optically Thick Clumps - Not the Solution to the Wolf-Rayet Wind Momentum Problem?}

\author{J. C. Brown \inst{1}, J.P. Cassinelli \inst{2}, Q. Li 
\inst{1,3}, A. F. Kholtygin \inst{4,6},  R. Ignace \inst{5}}

\institute{ Department of Physics and Astronomy, University of Glasgow,
Glasgow, G12 8QQ, UK
\and
  Department of Astronomy, University of Wisconsin-Madison, USA
\and
Department of Astronomy, Beijing Normal University, China
\and
Astronomical Institute, St.~Petersburg University, Saint Petersburg State University, V.V.Sobolev Astronomical Institute, 198504 Russia
\and
Department of Physics, Astronomy, \& Geology, East Tennessee State University, USA
\and
Isaac Newton Institute of Chile, St.Petersburg Branch
}
\offprints {john@astro.gla.ac.uk(JCB); li@astro.gla.ac.uk(QL)}
\authorrunning{Brown,  et~al.}

\titlerunning {Wind momentum problem}
%\maketitle

\date{}
\abstract {The hot star wind momentum problem $\eta=\mdot
\varv_{\infty}/(L/c) \gg 1$ is  revisited, and it is shown that the
conventional belief, that it can be solved by a combination of  clumping
of the  wind  and multiple scattering of photons, is not self-consistent
for optically thick clumps. Clumping does reduce the mass  loss rate
$\mdot$, and hence the momentum supply, required to generate a specified
radio emission measure  $\varepsilon$, while multiple scattering
increases the delivery of momentum from a specified stellar  luminosity
$L$. However, in the case of thick clumps, when combined the two
effects act in opposition rather than in unison since clumping  reduces
multiple scattering. From basic geometric considerations, it is shown
that this reduction in momentum delivery by  clumping more than offsets
the reduction in momentum required, for  a specified $\varepsilon$. Thus
the ratio of momentum deliverable to momentum required is maximal for a
smooth wind and the momentum problem remains for the thick clump case.
In the case of thin clumps, all of the benefit of clumping in reducing
$\eta$ lies in reducing $\mdot$ for a given $\varepsilon$ so that
extremely small filling factors $f\approx 10^{-4}$ are needed.

It is also shown that clumping affects the inference of $\mdot$ from radio
$\varepsilon$ not only by changing the emission measure per unit mass but
also by changing the radio optical depth unity radius $R_{\rm {rad}}$,
and hence the observed wind volume, at radio wavelengths. In fact, for
free-free opacity $\propto n^2$, contrary to intuition, $R_{\rm {rad}}$
increases with increasing clumpiness.

}
\maketitle

%\newpage
%\tableofcontents
%\vskip0.5cm

\section {Momentum problem}

If one infers the mass loss rate $\mdot$ for hot  massive
(especially Wolf-Rayet) stars from the radio emission measure
$\varepsilon=\int_Vn^2dV$, using a smooth spherical wind model, one
finds that the wind `momentum' rate  $\dot p=\mdot\varv_{\infty} =\eta
\, L/c$ involves $\eta\gg 1$, where $L/c$ is the radiative momentum
outflow rate (Cassinelli \& Castor \cite{cassinelli}).  Insofar as such
winds are believed to be radiatively driven, this poses a `momentum
problem', the solution of which has long been a hot topic in the field
(Barlow et~al. \cite{barlow}; Abbott et~al. \cite{abbott}; Cassinelli
\& van der Hucht \cite{cassinelli87}; Willis \cite{willis}; Lucy \&
Abbbot \cite{lucy}; Springmann \cite{springmann94}; Springmann \& Puls
\cite{springmann95}; Gayley, Owocki, \& Cranmer  \cite{gayley}; Owocki \&
Gayley \cite{owocki99}). Estimates of $\eta$ vary according to assumptions
(e.g., arguing for a high value of $L$) but values of $\eta$ ranging
up to nearly 100 are mentioned (Hamann \& Koesterke \cite{hamann}).
There are two main strands of argument quite widely believed to combine
to solve the momentum problem, one being mainly observationally driven
and the other mainly theoretical.

The values of $\dot M$ associated with these large $\eta$ are those
inferred from a smooth spherical wind density model, the radio emitting
material filling the volume. The contribution to $\varepsilon$  from
any volume element $\Delta V$ is $\Delta \varepsilon \approx n^2\Delta
V\propto \dot M^2/\Delta V$.  If, however, the material is clumpy,
filling only a fraction $f=\langle n\rangle^2/\langle n^2\rangle$
of the volume, then $\Delta\varepsilon$ is enhanced by a factor $1/f$
for a given $\dot M$. The mass loss rate $\dot M$ required to generate
an observed $\varepsilon$ thus scales as $\dot M \propto f^{1/2}$ in
clumpy winds. For strong clumping ($f \ll 1$), this ameliorates the
momentum problem, though the $f=10^{-4}$ required to reduce $\dot
M\varv_{\infty}$ by a factor of 100 seems very unlikely, so this
clumping effect alone cannot be the complete answer (e.g., Nugis \&
Lamers \cite{nugis}, cite clumping corrected mass-loss rates yielding
$\eta \approx 6$).  For example, making clumps very small increases their
radio optical thickness and may make optically thin emission measures
irrelevant. There is extensive observational evidence for large scale
clumping in WR winds: the presence of narrow emission features moving
out on broad wind emission lines (e.g., Robert et~al. \cite{robert89};
Robert et~al. \cite{robert91}; Moffat \& Robert \cite{moffat}; Kholtygin
\cite{kholtygin}); broad band photometric and polarimetric fluctuations
(e.g., Brown et~al. \cite{brown}; Li et~al. \cite{li}); and the absence
of strong electron scattering wings (which scales as $\langle n\rangle^2$
rather than $\langle n^2\rangle$ (Hillier \cite{hillier91})).

On the theory side, it has long been recognised that the limit $\dot
p=\dot M\varv_{\infty}\approx L/c$ is only true if (all) photons are
scattered once only. If the wind scattering optical depth is high, the
photons can, loosely speaking, be scattered `back and forth' across the
wind delivering momentum of up to $2h\nu/c$ at each scattering (for thick
clumps) until $\nu$ is progressively dissipated by Doppler reddening at
each momentum-delivering scattering on the moving matter. The nature of
this multiple scattering has been described with progressively greater
insight over the years. In particular, Gayley \etal\ (\cite{gayley})
showed that scattering back and forth across the entire wind is not
required. Instead, the momentum is delivered in a series of random
semi-local scatterings of photons as they diffuse outward, provided
successive scatterings involve long enough paths to  sample different
matter velocities. The essential feature is that of the large scattering
optical depth  $\tau$, which enhances the momentum delivery rate to $\tau
L/c$ (e.g., Friend \& Castor \cite{friend}; Kato \& Iben \cite{kato};
Netzer \& Elitzur \cite{netzer}; Gayley \etal\ \cite{gayley}), because
the diffusive delivery scales with the number of scatterings $N_s$
as $N_s^{1/2}$, while $N_s=\tau^2$.  Since the predominant driver
is via the large opacity/cross-section associated with lines, Gayley
et~al. (\cite{gayley}); and Owocki \& Gayley (\cite{owocki99}) have
suggested that the issue is not so much a momentum problem as an opacity
problem.

The massive WR winds are still believed to be driven by line opacity
(e.g., Lucy \& Abbott \cite{lucy}).  Unlike the less massive winds of
OB stars, the WR winds have significant ionization gradients, that can
substantially alter the line opacity distribution with radius in the flow
(Herald \etal\ \cite{herald}; Vink, de Koter, \& Lamers \cite{vink}).
Consequently, as the photons move away from the star, interact with
a certain line opacity that exists at some radius $r$ in the flow, and
then escape, the photons encounter a new line opacity distribution
at a different radius.  Consequently, if there are gaps in the line
distribution at one radius, those gaps can be filled by a different line
distribution that exists in another part of the wind flow.  The opacity
problem then represents how effectively all of these gaps are ``filled''.

Photon escape at gaps in the line frequency forest reduces the flux
mean opacity (or flux mean cross section $\sigma$  per particle in our
formulation) used in the gray approximation. The maximum that can be
achieved by multiple scattering is reached when the number of random
scatterings is so great as to Doppler shift the photons down to near
zero frequency, the maximum  Doppler shift per scattering being of
order $ \varv/c$ for wind speed $\varv$. This requires $\tau\approx
N_s^{1/2}\approx c/\varv$ ($\approx 100$), implying $\dot M \varv_{\infty}\approx
L/c\times c/\varv_{\infty}$ or $\dot M \varv_{\infty}^2/2\approx L/2$ 
which is the energy
conservation limit. Available calculations of multiple scattering with
real opacities can yield $\eta$ gains of order 10, that may explain
some WR~winds, but not the more extreme cases in which $\eta \sim 10^2$
is required.

Since reduction of $\dot M$ (for a given $\varepsilon$) by clumping
and increase of momentum delivery by multiple scattering can each
offer a factor of order 10 reduction in the momentum problem, there
seems to be growing widespread belief that the momentum problem can
be laid to rest (e.g., Conti \cite{conti}). However, this involves the
tacit assumption that these two factors can operate {\em independently}
and {\em constructively}, the impact of clumping on the effectiveness
of multiple scattering never having been addressed (Hillier \& Miller
\cite{hillier99}) (although Shaviv (\cite{shaviv}) has discussed the
related topic of how optically thick clumps increase the Eddington luminosity).
Here we show, using simple geometric
arguments, that this assumption is incorrect in the case of optically
thick clumps, and that clumping, while reducing $\dot M$, also reduces
$\tau$, so making multiple scattering less effective. Essentially
this is because clumping reduces the number of scattering centres
compared to scattering off of atoms and also reduces $<n>$, for a given
$\varepsilon$. (Note that when discussing the effects of clumping it is
essential to keep in mind that the observed $\varepsilon$ is held fixed.
This fact is sometimes overlooked).

 We find quantitatively that, for thick clumps, the
reduction in multiple scattering  momentum delivery more than offsets
the reduction in momentum required, the nett effect being that clumping
worsens the momentum problem rather than solving it.

\section{Single clump} 

To illustrate the point, we first consider one thick
scattering clump of mass $M$ composed of atoms/ions of mass $m$. This
is taken to have very high internal optical depth in the line-driving
wavelength range so the clump as a whole is the scattering centre.
 Since we are not concerned with the wind speed profile $\varv(r)$
but only with the final wind speed and momentum, we here approximate
clumps as moving radially with speed $\varv\approx \varv_\infty$
and to have the shape of a conical slice of radial thickness
$\delta$ and solid angle $\Omega$, the volume of the cone being
$r^2\Omega \delta$ at distance $r$. We assume the clump to be
optically thin at radio wavelengths, so its radio flux depends
on the emission measure, but optically thick to lines for the stellar
radiation at short wavelengths that are
responsible for driving the flow. 

The emission measure
$\varepsilon_1=n^2V$ (which measures the radio emission rate) of a
single clump is 

\begin{equation} 
\varepsilon_1=\frac{(M/m)^2}{r^2\Omega
\delta}=\frac{\varepsilon_{1o}}{x^2}           
\label{em1} 
\end {equation}

\noindent where $\varepsilon_{1o}=(M/m)^2/(R_{\rm rad}^2\Omega\delta)$
the clump emission measure at $r=xR_{\rm rad}=R_{\rm rad}$,
for $R_{\rm rad}$ is the radius of  the radio photosphere which may be
hundreds of times larger than the optical photosphere radius. Note that,
for a prescribed $\varepsilon_1$, $M\propto\sqrt{\Omega}$ for any chosen
$r,\delta$.

On the other hand, the available rate of delivery of momentum is 
\begin {equation}
\pdot_{1,avail}=\frac{L}{c}\frac{\Omega}{4\pi}. \label {pavail1} 
\end {equation}
where we ignore scale factors of order unity due to the effects of gravity and of
the backward scattering angular distribution function.  The rate of momentum delivery
required is 
\begin{equation} 
\dot p_{1,req}\approx\frac{M \varv}{r/\varv}.
\end{equation} 
It follows that, for a given $\varepsilon_1$, the
effectiveness of momentum delivery
to a single clump is
\begin {equation} 
\Psi_1=\frac{\dot
p_{1,avail}}{\dot p_{1,req}}=\left[\frac{L}{4\pi m
\varv^2c\varepsilon_1^{1/2}}\right]\left(\frac{\Omega}{\delta}\right)^{1/2}
\label{psi1} 
\end{equation} 
This decreases as we make $\Omega$ smaller
-- i.e., as we make the clump clumpier -- because the momentum $\dot
p_{1,req}$ required $\propto M\propto\sqrt{\Omega}$, but the momentum
$\dot p_{1,avail}$ available $\propto\Omega$, and the decline of the
latter with $\Omega$ is dominant for a single clump. That is, making
the $\Omega$ of a single clump smaller does reduce $\dot p_{1,req}$
for a given $\varepsilon_1$ but reduces $\pdot_{1,avail}$ even more. So
shrinking one clump of a given $\varepsilon_1$ makes it harder to drive
it to terminal speed of known value $\varv_{\infty}$.

Compressing the clump radially does help (in this single thick clump case)
since reducing $\delta$ reduces $M\varv$ for prescribed $\varepsilon_{1},
\Omega$.

\section {Multiple clumps}

We now have to consider the effect of multiple scattering in the case
of a multiple clump wind, since multiple scattering cannot occur in
the case of an individual discrete clump. In doing so we take all
clumps to be optically thick in the UV but thin in radio,
identical in size and mass, and use the  gray opacity
approximation, the clumps being driven by a spectral mean `continuum'
radiation flux.  We are of course well aware that in reality there will
be a distribution of clump sizes and masses. However, if one can prove
that for any  specific clump parameters, clumping reduces the benefit
of multiple scattering, then the same must be true of the sum over any
distribution of clump parameters so long as they remain thick.  
Put another way, the arguments that
clumping a wind increases its emission measure, that multiple scattering
increase momentum delivery, and that clumping reduces multiple scattering
all derive essentially from geometric arguments and have nothing to do with
the details of opacity or of clump size distribution (other than being thick).

Retention of the conical slice shape described above, taking $\Omega$
and $\delta$ independent of $r$, means that the clumps expand in 2-D
(transversely) rather than in 3-D, which is reasonable for a highly
supersonic wind. The constant $\Omega$, $\delta$ assumption also
means that, for constant $\varv$, clumps occupy the same fraction
(constant filling factor $f$) of the volume at all $r$. For spherical
(3-D) clump expansion, linear radial expansion ($\delta \propto r$)
would result, for constant $\varv$, in radial merging of clumps, which
corresponds to an $r$--dependent filling factor $f$ with $f\to1$ as 
clumps merge. Situations with non-constant filling factor $f=f(r)$
have been discussed by Nugis, Crowther, \& Willis (\cite{nugis98});
Hillier \& Miller (\cite{hillier99}); and Ignace et al. (\cite{ignace03}).
We assume clumps are, on average, emitted uniformly over the stellar
surface at a rate $\Cdot$ in clumps per second. Then the space density
of clumps at $r$ is

\begin{equation}
\nf(r)=\frac{\Cdot}{4\pi r^2 \varv},  \label{nfdef}
\end{equation}
where we again approximate $\varv=$ constant $=\varv_{\infty}$ and the radio
emission measure of one clump is again given by Eq. (\ref {em1}).
Using Eqs.~(\ref{em1}) and (\ref{nfdef}), the total emission measure
can be written

\begin{eqnarray}
\varepsilon=\int^{\infty}_{R_{\rm rad}} 4\pi r^2dr {\cal N} \varepsilon_1
=\frac{\Cdot}{R_{\rm rad}\varv} \left(\frac{M}{m}\right)^2\frac{1}
{\Omega\delta}=\frac{\dot C R_{\rm rad}}{\varv}\varepsilon_{1o}=N_{eff} 
\varepsilon_{1o},    \label{epsilon}
\end {eqnarray}
The last form is interesting, showing that the total emission measure
$\varepsilon$ is just the initial emission measure of one clump
$\varepsilon_{1o}$ at $r=R_{\rm rad}$ times an effective number of clumps
$N_{eff}= \dot C R_{\rm rad}/\varv$, namely that located in the
range $R_{\rm rad}\le r < 2R_{\rm rad}$.

The  mass loss rate $\mdot$ and the momentum delivery rate $\dot p_{req}$ required are
\begin{equation}
\mdot=\Cdot M           \label{massloss}
\end{equation}
and
\begin{equation}
\dot p_{req}=\Cdot M \varv.             \label {pdot}
\end{equation}
where we neglect scale factors of order unity as we did in Eq.(\ref{pavail1}).
By Eqs. (\ref{epsilon}) and (\ref{pdot}), we get the momentum 
delivery rate required for a given total wind emission
measure $\varepsilon$ as a function of $M$, $\Omega$, namely
\begin{equation}
\dot p_{req}=m^2 \varv^2 R_{\rm rad}  \varepsilon \frac {\Omega\delta}{M}.   \label{pre1}
\end{equation}
We want to compare this with the momentum delivery rate available 
from multiple scattering of stellar photons and we take this to
be given by (cf. Section 1) 
\begin{equation}
\pdot_{avail}=\frac{L}{c}N_s^{1/2}=\frac{L}{c}\tau     \label{pav1}
\end{equation}
where $\tau$ is the mean (gray approximation) line  scattering optical
depth of the wind and $N_s=\tau^2$ is the number of scatterings of an
escaping photon.  $\tau$ is also the `covering factor' or the total
solid angle of all the clumps as seen from the star divided by $4\pi$
- see Appendix.

The wind optical depth for starlight due to lines treated in the
gray approximation is (for individually thick clumps)
\begin{equation}
\tau=\int^{\infty}_{R_{\rm UV}} nQ dr
\end{equation}
where $n$ is the density of scatterers, $Q$ is the cross 
section of scatterers, and $R_{\rm UV}$ is the UV photosphere associated with
the dominant line driving. 

We choose to split the range into two
sectors, $r<d$ and $r>d$, where $d$ is the distance at which 
an individual clump becomes optically thin radially. At
$r<d$ the individual $scatterer$ is a clump of area $r^2\Omega$ 
and thickness $\delta$, while at $r>d$, it is an ion of
area
$\sigma$ (the actual value adopted for $\sigma$ being some frequency 
average over lines).
Thus the optical depth integral expands to
\begin{equation}
\tau=\int^{d}_{R_{\rm UV}}\nf (r) r^2 \Omega dr + \int^{\infty}_{d} 
\nf(r)\frac{M}{m}\sigma dr
=\frac{\Cdot R_{\rm UV}}{4\pi \varv}\left[\Omega\left(\frac{d}{R_{\rm UV}}-1\right)+\frac 
{M\sigma}{mR_{\rm UV}^2}\frac{R_{\rm UV}}{d}\right]     \label{taua}
\end{equation}
where $d$ satisfies 
\begin{equation}
\frac{M\sigma}{m}\frac{1}{\Omega d^2}=1           \label{condit}
\end{equation}
and thus
\begin{equation}
\frac{d}{R_{\rm UV}}=\left(\frac {M\sigma}{mR_{\rm UV}^2}\right)^{1/2}\frac{1}{\Omega^{1/2}}.
\end{equation}
Then equation (\ref{taua}) becomes
\begin{equation}
\tau=\frac{\Cdot R_{\rm UV}}{4\pi \varv}\left[2\left(\frac{M\sigma}{mR_{\rm UV}^2}\right)^{1/2}\Omega^{1/2}-\Omega\right]
=\frac{\Cdot}{4\pi 
\varv}\left(\frac{M\sigma}{m}\right)^{1/2}\Omega^{1/2}\left[2-\left(\frac{mR_{\rm UV}^2}{M\sigma}\right)^{1/2}\Omega^{1/2}\right]. 
     \label{taub}
\end{equation}

Consider the second term in expression (\ref{taub}). The ratio
$\frac{M\sigma}{m}$ is the total area of all the atoms in a clump, and
$R_{\rm UV}^2\Omega$ is the total area of a clump at $r=R_{\rm UV}$. Since our analysis
deals with individually thick clumps, we require $\frac {M\sigma}{m}\gg
R_{\rm UV}^2\Omega$, so we can get $d\gg R_{\rm UV}$ and neglect the second term
in Eq. (\ref{taub}) and write

\begin{equation}
\tau=\frac{\Cdot }{2\pi \varv}\left(\frac{M\sigma}{m}\right)^{1/2}\Omega^{1/2}. 
\label {tauc}
\end{equation}

If we express $\Cdot$ in terms of $\varepsilon$ by using Eq. 
(\ref{epsilon}), then Eq. (\ref{tauc}) becomes

\begin{equation}
\tau=\frac{ R_{\rm rad}\,\sigma^{1/2}}{2\pi}\left(\frac{m}{M}\right)^{3/2} 
\varepsilon\,\delta \,\Omega^{3/2}.         \label {tau}
\end{equation}

Now we get the available momentum using Eqs. (\ref{pav1}) and
(\ref{tau}) in terms of $M$, $\Omega$ for a given $\varepsilon$, namely

\begin{equation}
\pdot_{avail}=\frac{L}{c}\left(\frac{R_{\rm rad}\sigma^{1/2}}{2\pi}\right) 
\left(\frac{m}{M}\right)^{3/2} \varepsilon \delta \Omega^{3/2}.     \label{pav2}
\end{equation}

Comparing Eqs. (\ref{pre1}) and (\ref{pav2}), we find a 
dimensionless measure of the effectiveness of momentum delivery,
as the ratio of momentum available to momentum required, as a 
function of $M$, $\Omega$, for a given $\varepsilon$, namely

\begin{equation}
\Psi=\frac{\pdot_{avail}}{\pdot_{req}}
=\left[\frac{L}{2\pi \varv^2cR_{\rm rad}}\left(\frac{\sigma}{m}\right)^{1/2}\right] 
\left(\frac{\Omega R_{\rm rad}^2}{M}\right)^{1/2} .    \label {psi}
\end{equation}
For a given star ($L$) and a wind of given speed $\varv$ and composition
($m,\sigma$), the expression in the bracket [ ] of Eq. (\ref{psi})
is constant, so we can  vary the value of $\Psi$  by varying the clump
parameter combination  $\Omega/M$.

The essential result is that $\Psi$ increases with increasing $\Omega/M$,
i.e., with increasing clump cross section per unit mass (which is
different from the single clump case of Eq. (\ref{psi1})). To
minimise the momentum problem (maximise $\Psi$) for a given mass $M$
(and thickness $\delta$), $\Omega$ should be as large as possible while
for a given $\Omega$ the mass $M$ should be as low as possible with,
in both cases $\dot C$ varying according to Eq. (\ref{epsilon}) to
ensure the correct $\varepsilon$. If we change (e.g., increase) $\delta$,
$\Psi$ does not change but $\dot C$ changes (falls) to maintain fixed
$\varepsilon$. Consequently, to maximise $\Psi$ we must make the clump
mass small, the clump angle large, and the clump thickness large with
correspondingly small $\dot C$, all of these corresponding to minimising
clumping.

It is also of interest to express $\Psi$ in terms of the volume filling
factor $f=\langle n\rangle^2/\langle n^2\rangle$ which can be expressed
(with $V_c=$ single clump volume $=\Omega r^2\delta$ at $r$)  as

\begin{equation}
f	=\frac{4\pi r^2 dr{\cal N}(r) V_c}{4\pi  r^2 dr}
	=\frac{\dot C R_{\rm rad}}{\varv}\frac{\Omega}{4\pi}\frac{\delta}{R_{\rm rad}}
	=N_{eff}f_\Omega f_r
\end{equation}
where $f_\Omega=\frac{\Omega}{4\pi}$ and $f_r=\frac{\delta}{R_{\rm rad}}$ are
solid angle and radial filling factors respectively. Alternatively,
the volume filling factor can be expressed as

\begin{equation}
f=\frac{R_{\rm rad}\varepsilon m^2}{4\pi}\left(\frac{\Omega \delta}{M}\right)^2.
\label{ff}
\end{equation}
Comparing Eqs. (\ref{psi} ) and (\ref{ff}), we see that, under our thick clump assumption,
$\Omega/M\propto f^{1/2}$ so that $\Psi \propto f^{1/4}$, so that  for
any values of ($\varepsilon,\delta$), decreasing $f$ \textit {decreases} the
effectiveness of momentum delivery.

All of the above shows  that, contrary to conventional `wisdom', in the case of thick
clumps, clumping does not help solve the momentum problem but actually makes
it worse.  

The case of a smooth wind can be considered a limit
of the clumpy case as the clumps blend.  
However, there are infinitely many clumpy cases that approach the smooth
case  as the clumps blend and it is easier to evaluate $\Psi=\Psi_o$
for the smooth case directly using $n_o(r)=\dot M/(4\pi r^2m\varv)$. Then
with subscript `$o$' denoting the smooth case, we get

\begin{equation}
\dot p_{o,req}=\dot M\varv,
  \label {pdoto}
\end{equation}

\begin{equation}
\varepsilon_o=\frac{\dot M^2}{m^2}\frac{1}{4\pi \varv^2R_{\rm rad}},
  \label {epsilono}
\end{equation}

\begin{equation}
\tau_o=\frac{\dot M\sigma}{4\pi m\varv R_{\rm UV}},
  \label {tauo}
\end{equation}
and
\begin{equation}
\dot p_{o,avail}=\frac{L}{c}\tau_o=\frac{L}{c}\frac{\dot 
M\sigma}{4\pi m\varv R_{\rm UV}}
  \label {pavailo}
\end{equation}
from which we deduce that

\begin{equation}
\Psi_o=\frac{L}{c}\frac{1}{4\pi \varv^2R_{\rm UV}}\frac{\sigma}{m} .
  \label {psio}
\end{equation}

Explicitly comparing the momentum delivery effectiveness for the clumpy
and smooth cases we have, by Eqs. (\ref {psi}) and (\ref {psio})

\begin{equation}
\frac{\Psi}{\Psi_o}=2\left[{\left(\frac{\Omega R_{\rm UV}^2}{M}\right)}
		/\left(\frac{\sigma}{m}\right)\right]^{1/2}
	=\left(\frac{(2R_{\rm UV})^2}{\frac{M\sigma}{m\Omega}}\right)^{1/2}
	=\frac{2R_{\rm UV}}{d}
	\label {Psi/Psio}
\end{equation}
which is clearly $\ll 1$ for clumps which are initially optically
thick. Note also that $\Psi$ involves the ratio of $\Omega (2R_{\rm UV})^2/M$ and
$\sigma/m$, respectively, the cross-sections per unit mass of clumps
and of atoms, while $\Psi_o$ involves only $\sigma/m$. Clearly, the
continuous limit corresponds to $\Omega (2R_{\rm UV})^2/M \to \sigma/m$ at $d
\to 2R_{\rm UV}$ (cf. Eq. (\ref{condit})), the `clumps' become individual
atoms, becoming thick at that point. Note that $d \rightarrow R_{\rm UV}$ is essentially
the limit where driving approaches the smooth wind limit, equally applicable to 
optically thin clumps.

Carrying this line of inquiry further, it is helpful to see how
Eq. (\ref{psi}) for $\Psi$ approaches the smooth limit $\Psi_o$.
We require that $\tau=\tau_o$ from Eqs. (\ref{tauc}) and (\ref{tauo}),
that $\varepsilon = \varepsilon_o$ from Eqs. (\ref{epsilon}) and
(\ref{epsilono}), and finally that $f=1$. These conditions are
met for

\begin{equation}
\frac{\dot C R_{\rm rad}}{\varv}\frac{\Omega}{4\pi}\frac{\delta}{R_{\rm rad}}
	=N_{eff}f_rf_\Omega =1,
\end{equation}

\begin{equation}
\frac{M\sigma}{m}=\Omega (2R_{\rm UV})^2,
                           \label{omega}
\end{equation}
and

\begin{equation}
\delta =\frac{16\pi m \varv R_{\rm UV}^2}{\mdot \sigma}
                                   \label{delta}
\end{equation}
which, on substitution in Eq. (\ref{psi}) gives $\Psi=\Psi_o$
(Eq. (\ref{psio})) as required.  To interpret Eq. (\ref{omega})
physically, note that $M\sigma/m$ is the total cross
sectional area of all the atoms in one clump while $\Omega(2R_{\rm UV})^2$
is the  cross sectional area of one clump at $r=2R_{\rm UV}$. These can
only be equal if clumps have the scale of individual atoms. Secondly,
Eq. (\ref{delta}) can be expressed as $4\pi(2R_{\rm UV})^2=\frac{\mdot \sigma\delta}{m\varv}$. 
Here $\frac{\mdot \sigma}{m\varv}$ is the total area 
of all clump atoms per unit radial distance, 
so $\frac{\mdot \sigma\delta}{m\varv}$ is the total area of all clump atoms 
in a scale length $\delta$. Thus, since $4\pi (2R_{\rm UV})^2$ is the spherical area 
at $r=2R_{\rm UV}$, the scale $\delta$ defines the range of $r$ around $2R_{\rm UV}$ 
in a smooth wind over which the clump atoms just cover the sphere.

\section {Discussion and conclusions}

\subsection{Conclusions regarding thick clumps}
Using a simple model we have shown that while clumping reduces the
mass loss rate of WR stars required by radio emission measures, it
also reduces the wind optical depth and hence multiple scattering and
momentum delivery. The nett result is that thick clumping worsens the momentum
discrepancy  rather than solving it. This is not the case for thin clumps, as we discuss
below.

\subsection{Discussion of assumptions}
First we comment on various simplifications we have made which might modify our
results somewhat. These include the approximation of constant $\varv$,
relaxation of which does not seem very likely to change our results
much since the radio emission measure is produced well out in the winds
beyond where wind acceleration starts. Secondly, in common with many authors, we have 
so far assumed that the inner boundary $R_{\rm rad}$ (and hence the volume) of the radio
emission measure region does not change with clumping. In fact, one might
expect, by analogy with the UV optical depth (Eq. (\ref{tau})), that
the radio optical depth might fall with increased clumping, reducing $R_{\rm rad}$
and increasing the radio source volume and emission measure. To check
this, we first want to know the radio optical depth $\tau'$ for a clumped
wind. This is roughly given by

\begin{equation}
\tau'(r)=\int_r^{\infty}{\cal N}(r)\frac{M}{m} \sigma' dr       \label{taupr}
\end{equation}
where $\sigma'$ is the relevant cross section per proton. However,
we have to note that the main radio absorption mechanism is free-free
opacity which is density dependent ($\sigma' \propto n_c \approx
1/r^2$) and we have to write

\begin{equation}
\sigma'=\sigma_o'\frac{n_c(r)}{n_o}                  \label{sigmap1}
\end{equation}
where $n_c=n_o$, and $\sigma'=\sigma_o'$ are defined in any reference level  $r=r_o$.
Then Eq.~(\ref{taupr}) becomes
\begin{equation}
\tau'(r)=\frac{\Cdot}{12\pi \varv}\frac{\sigma_o'}{n_o}\left(\frac {M}{m}\right)^2\frac{1}{\Omega \delta r^3},       \label{sigmap2}
\end{equation}
so that now $R_{\rm rad}$ is given by $\tau'(R_{\rm rad})=1$, namely
\begin{equation}
R_{\rm rad}=\left(\frac{\mdot\,\sigma_o'}{12\pi \varv\,n_o m^2}\right)^{1/3}\,\left( \frac{M}{\Omega \delta}\right)^{1/3}.                                                           \label{radior}
\end{equation}
Consequently, increasing $M$ for given $\Omega$, $\delta$ or decreasing
$\Omega$, $\delta$ for a given $M$ (i.e., increasing clumpiness) actually
makes $R_{\rm rad}$  bigger, not smaller, in a clumpy wind because of the $density$
dependence of free-free absorption.

The corresponding emission measure expression is now as before but based on the new clumping dependent value 
in Eq.~(\ref{radior}) of $R_{\rm rad}$ which leads to
\begin{equation}
\varepsilon=\frac{3^{1/3}}{(4\pi)^{2/3}}\left(\frac{\mdot}{\varv}
\right)^{2/3} \left(\frac {n_o}{m^4\sigma_o'} \right)^{1/3}\left(\frac{M}{\Omega \delta}\right)^{2/3}                    \label{radioep}
\end{equation}
which does increase as we increase $M/\Omega \delta$ (i.e.,
clumpiness) for a given $\mdot$ but now with $\varepsilon\propto [\mdot
M/(\Omega\delta)]^{2/3}$, instead of $[\mdot M/(\Omega\delta)]$ for the
constant $R_{\rm rad}$ case. Thus although clumpiness still reduces
$\mdot$ for a given $\varepsilon$, it does so less than with constant
$R_{\rm rad}$ and likewise, thick clumps are now even less helpful to
the momentum problem.

\subsection{Discussion of thin clumps}
Throughout, we have considered here the case when each clump is an optically thick
scatterer at line-driven wavelengths.  The other case of relevance to reducing $\eta$
is where there is strong clumping ($f \ll 1$) but with very large numbers of very small scale clumps, each
optically thin. We have considered here the case when each clump is an optically thick
scatterer at line driving wavelengths but with a large overall wind optical depth. In this case
our analysis does not really apply and the driving has to be described by multiple scatterings
over the small Sobolev optical depths of many successive individual thin scatterers. The nett effect here
is to yield the same momentum delivery rate as in a smooth wind but to reduce $\mdot$ for a given $\varepsilon$
by increasing the optically thin radio emission measure $\varepsilon_1$ of each clump. In this thin clump limit,
clumping does not reduce momentum delivery by photon escape, as happens for thick clumps, but reduces $\mdot$
for a given $\varepsilon$. As a solution to the momentum problem, this scenario puts the entire onus on the 
reduction of $\mdot/\varepsilon$ and of $\eta$ by a factor $f^{1/2}$ and for the extreme case of $\eta \approx 100$
requires $f\approx 10^{-4}$. How such a huge compression of wind matter into clumps is achievable physically, and whether
it can be done without making the clumps optically thick in the radio (bearing in mind that the radio opacity $\propto n^2$)
are questions that must be addressed before clumping solutions of the momentum problem, are to be accepted.

\subsection{Other solutions to the momentum problem?}
Since clumping solutions, thick or thin, remain questionable, we briefly mention
here some other possibly relevant factors.  Some have advocated non-spherical
models for the massive WR flows.  Lamers \& Pauldrach (\cite{lamers})
developed a bi-stability model for early-type stars (e.g., B[e] stars),
and Poe \etal\ (\cite{poe}) proposed a two-component model with fast winds
from polar regions and a slow flow from equatorial regions, later termed
the Luminous Magnetic Rotator model (Cassinelli (\cite{cassinelli91})).
In aspherical models such as these, the high radio flux arises from
the denser equatorial region, whereas the high terminal speeds derive
from a line-of-sight that lies perhaps in a broad polar region. These
two-component structures have not been widely accepted because few WR
stars show substantial (non-varying) intrinsic polarizations, as would
be expected for stars that have a dense equatorial flow  (Harries,
Hillier, \& Howarth \cite{harries}). However, Taylor \& Cassinelli (1992)
studied the cancellation of polarization owing to  a more tenuous polar wind,
and it was surprisingly effective.  Whether or not the solution to the momentum problem
lies in the aspherical structure of such a rotational model, or yet some other aspherical picture, our
paper has shown that the wide spread belief that the solution lies in the clumpy $+$
multiple scattering picture is too simplistic.

\begin {acknowledgements}

The authors acknowledge support for this work from: a NATO Collaboration
Grant (AFK, JCB, JPC); a UK PPARC Research Grant (JCB); NASA Grant Number
TM4--5001X (JPC, JCB); Royal Society Sino-British Fellowship Trust Award (QL);
a NSFC grant 10273002 (QL); and a RFBR grant 01-02-16858 (AFK).
Thanks to the referee (Ken Gayley) whose comments
led to a significant improvement of the paper.

\end{acknowledgements}

\appendix

\section{Relation between Covering factor $Y$, number of scatterings
$N_s$, and optical depth $\tau$}

$Y$ is the fraction of the solid angle around a star that is covered 
by scatterers. Let $A$ and $\omega$ be the cross section and solid angle 
for one scatterer at $r$, so that $\omega=A(r)/r^2$ and let $\nf_s(r)$ be the space density of 
scatterers, then the covering factor at $r$ is the total solid angle of all the scatterers divided by $4\pi$, namely
\begin{equation}
Y=\frac{1}{4\pi}\int^{\infty}_r \nf_s(r)\, \omega\,4\pi \,r^2\,dr 
=\frac{1}{4\pi}\int^{\infty}_r \nf_s(r) \,\frac{A(r)}{r^2}\,4\pi \,r^2\,dr
=\int^{\infty}_r \nf_s(r) \,A(r) \,dr =\tau(r).
\end {equation}

A photon travelling in a medium with typical size $D$, density $\nf_s$, 
and particle cross section $A$, on average
undergoes $N_s$ scatterings before escaping.  Its  mean free path 
$l=1/(\nf_s A)$, is related to $N_s$ by $l\sqrt{N_s}=D$, so
\begin{equation}
N_s=\left(\frac {D}{l}\right)^2=(\nf_s A\,D)^2 =\tau^2.
\end{equation}


\begin{thebibliography}{}

\bibitem[1986]{abbott} Abbott, D. C., Torres, A. V., Bieging, J. H., \&
Churchwell, E., 1986, ApJ, 303, 239

\bibitem[1981]{barlow} Barlow, M. J., Smith, L. J., \& Willis, A. J.,
1981, MNRAS, 196, 101

\bibitem[1995]{brown} Brown, J. C., Richardson, L. L., Antokhin, I., Robert, C., Moffat, A. F. J., \& St-Louis, N., 1995, A\&A 295, 725

\bibitem[1973]{cassinelli} Cassinelli, J.~P., \& Castor, J.~I., 1973,
ApJ, 179, 189

\bibitem[1987]{cassinelli87} Cassinelli, J.~P., \& van der Hucht, K. A.,
1987, in Instabilities in Luminous Early Type Stars,  ed. H. J. G. Lamers
\& C. W.H. De Loore, Astrophysics and Space Science Library (Dordrecht:
Reidel), vol. 136,  231

\bibitem[1991]{cassinelli91} Cassinelli, J.~P., 1991, in Wolf-Rayet Stars
and Interrelations with Other Massive Stars in Galaxies: Proc. IAU
Symp. 143, ed. K. A. van der Hucht \& B. Hidayat.,  (Dordrecht:
Kluwer), 289

\bibitem[1995]{conti} Conti, P. S., 1995, in Wolf-Rayet Stars: Binaries;
Colliding Winds; Evolution: Proc. IAU Symp. 163, ed. K. A. van der Hucht
\& P. M. Williams,  (Dordrecht: Kluwer), 565

\bibitem[1983]{friend} Friend, D.~B., \& Castor, J.~I., 1983, ApJ,
272, 259

\bibitem[1995]{gayley} Gayley, K. G.,  Owocki, S. P., \& Cranmer, 1995,
ApJ, 442, 296

\bibitem[1998]{hamann} Hamann, W. R., \& Koesterke, L.
1998, A\&A, 333, 251

\bibitem[1998]{harries} Harries, T.~J., Hillier, D.~J., \& Howarth, I.~D.,
1998, MNRAS, 296, 1072

\bibitem[2000]{herald} Herald, J.~E., Schulte-Ladbeck, R.~E., 
Eenens, P.~R.~J., \& Morris, P., 2000, ApJS, 126, 469

\bibitem[1991]{hillier91} Hillier, D. I., 1991, A\&A, 247, 455

\bibitem[1999]{hillier99} Hillier, D.~ I., \& Miller, D.~L., 1999, ApJ, 519, 354

\bibitem[1992]{hucht} van der Hucht, K.~A., 1992, A\&AR, 4, 123

\bibitem[2000]{ignace} Ignace, R., Oskinova, L. M., Foullon, C., 2000,
MNRAS, 318, 214

\bibitem[2003]{ignace03} Ignace, R., Quigley, M.~F., \& Cassinelli, J.~P.
2003, Apj, 596, 538

\bibitem[1992]{kato} Kato, M., \& Iben, I., 1992, ApJ, 394, 305

\bibitem[1995]{kholtygin} Kholtigin, A. F., 1995, in Wolf-Rayet Stars:
Binaries; Colliding Winds; Evolution: Proc. IAU Symp. 163, ed. K. A. van
der Hucht \& P. M. Williams,  (Dordrecht: Kluwer), 160

\bibitem[1991]{lamers} Lamers, H.~J.~G.~L.~M., \& Pauldrach, A.~W.~A.,
1991, A\&A 244, L5

\bibitem[2000]{li} Li, Q., Brown, J.~C.,  Ignace, R., Cassinelli, J.~P.,
\& Oskinova, L.~M., 2000, A\&A 357, 233

\bibitem[1993]{lucy} Lucy, L. B., \& Abbott, D. C., 1993, ApJ, 405, 738

\bibitem[1991]{moffat} Moffat, A.~ F.~ J., \& Robert, C., 1991, in
Wolf-Rayet Stars and Interrelations with Other Massive Stars in Galaxies:
Proc. IAU Symp. 143, ed. K. A. van der Hucht \& B. Hidayat.,  (Dordrecht:
Kluwer), 109

\bibitem[1993]{netzer} Netzer, N., \& Elitzur, M., 1993, ApJ, 410, 701

\bibitem[1998]{nugis98} Nugis, T., Crowther, P.~A., \& Willis, A.~J., 1998,
A\&A, 333, 956

\bibitem[2000]{nugis} Nugis, T., \& Lamers, H.~J.~G.~L.~M., 1993, ApJ, 410, 701

\bibitem[1995]{owocki95} Owocki, S. P., \& Gayley, K. G., 1995, in
Wolf-Rayet Stars: Binaries; Colliding Winds; Evolution: Proc. IAU
Symp. 163, ed. K. A. van der Hucht \& P. M. Williams,  (Dordrecht:
Kluwer), 138

\bibitem[1999]{owocki99} Owocki, S. P., \& Gayley, K. G., 1999, in
Wolf-Rayet Phenomena in Massive Stars and Starburst Galaxies: Proc. IAU
Symp. 193, ed. K. A. van der Hucht, G. Koenigsberger, \& P. R. J. Eenens,
ASP (Francisco, Calif), 157

\bibitem[1989]{poe} Poe, C.~H., Friend, D.~B., \& Cassinelli, J.~P., 1989, ApJ, 337, 888

\bibitem[1989]{robert89} Robert, C.,  Moffat, A.~F. ~J., Bastien, P.,
Drissen, L., \& St.-Louis, N., 1989, ApJ, 347, 1034

\bibitem[1991]{robert91} Robert, C., Moffat, A. F. J., \& Seggewiss, W.,
1991, in Wolf-Rayet Stars and Interrelations with Other Massive Stars in
Galaxies: Proc. IAU Symp. 143, ed. K. A. van der Hucht \& B. Hidayat.,
(Dordrecht: Kluwer), 147

\bibitem[1998]{shaviv} Shaviv, N. J., 1998, ApJ, 494, L193

\bibitem[1994]{springmann94} Springmann, U., 1994, A\&A, 289, 505

\bibitem[1995]{springmann95} Springmann, U., \& Puls, J., 1995, in
Wolf-Rayet Stars: Binaries; Colliding Winds; Evolution: Proc. IAU
Symp. 163, ed. K. A. van der Hucht \& P. M. Williams,  (Dordrecht:
Kluwer), 170

\bibitem[2000]{vink} Vink, J.~S., de Koter, A., \& Lamers, H.~J.~G.~L.~M.,
2000, A\&A, 362, 295

\bibitem[1991]{willis} Willis, A.~J., 1991, in Wolf-Rayet Stars and
Interrelations with Other Massive Stars in Galaxies: Proc. IAU Symp. 143,
ed. K. A. van der Hucht \& B. Hidayat.,  (Dordrecht: Kluwer), 265

\end{thebibliography}
\end {document}